\documentclass[%
 reprint,
 superscriptaddress,
 showpacs,
 amsmath,amssymb,
 aps,
]{revtex4-1}

\usepackage{graphicx}%
\usepackage{dcolumn}
\usepackage{bm}

\def\U#1{{%
\def\O{\mbox{O}}
\def\u{\mbox{u}}
\mathcode`\u=\mu
\mathcode`\O=\Omega
\mathrm{#1}}}

\def\ii{{\mathrm{i}}}
\def\ee{{\mathrm{e}}}
\def\dd{{\mathrm{d}}}
\def\Re{\mathop{\mathrm{Re}}}
\def\Im{\mathop{\mathrm{Im}}}

\def\bra#1{\langle #1|}
\def\ket#1{|#1\rangle}
\def\bracket#1{\langle#1\rangle}

\def\sub#1{_{\mathrm{#1}}}
\def\sur#1{^{\mathrm{#1}}}
\def\vct#1{{\mathchoice{\mbox{\boldmath$#1$}}{\mbox{\boldmath$#1$}}%
  {\mbox{\scriptsize\boldmath$#1$}}{\mbox{\scriptsize\boldmath$#1$}}}}

\newcommand{\nn}{\nonumber \\}
\newcommand{\Tr}{\mathrm{Tr}}

\begin{document}

\title{Simulating the classical XY model with a laser network}

\author{Shuhei Tamate}
\affiliation{
National Institute of Informatics, Hitotsubashi 2-1-2, Chiyoda-ku, Tokyo 101-8403, Japan
}
\email{tamate@nii.ac.jp}
\author{Yoshihisa Yamamoto}
\affiliation{%
ImPACT program, The Japan Science and Technology Agency, Gobancho 7, Chiyoda-ku, Tokyo 102-0076, Japan
}
\affiliation{%
E. L. Ginzton Laboratory, Stanford University, Stanford, CA94305, USA
}%
\author{Alireza Marandi}
\affiliation{%
E. L. Ginzton Laboratory, Stanford University, Stanford, CA94305, USA
}
\author{Peter McMahon}
\affiliation{%
E. L. Ginzton Laboratory, Stanford University, Stanford, CA94305, USA
}
\author{Shoko Utsunomiya}
\affiliation{
National Institute of Informatics, Hitotsubashi 2-1-2, Chiyoda-ku, Tokyo 101-8403, Japan
}

\date{\today}

\begin{abstract}
Drawing fair samples from the Boltzmann distribution of a statistical model is
a challenging task for modern digital computers.
We propose a physical implementation of a Boltzmann sampler for the classical XY model by using a laser network.
The XY spins are mapped onto the phases of multiple laser pulses in a fiber ring cavity
and the steady-state distribution of phases naturally realizes the Boltzmann distribution of the corresponding XY model.
We experimentally implement the laser network by using an actively mode-locked fiber laser
with optical delay lines, and demonstrate Boltzmann sampling for a one-dimensional XY ring.
\end{abstract}

\maketitle

Sampling from the Boltzmann distribution of statistical models is
one of the key techniques to understand the physics of many-body systems.
In recent years, due to the great success of restricted Boltzmann machines \cite{Smolensky1986} for
various tasks in machine learning,
Boltzmann sampling has attracted great attention in the field of computer science.

The conventional way to sample from the Boltzmann distribution is
based on Markov chain Monte Carlo (MCMC) procedures.
However, drawing fair samples from a given Hamiltonian is a computationally difficult task (exact sampling is NP-hard \cite{Barahona1982} and approximate sampling is hard unless $\mathrm{RP} \neq \mathrm{NP}$ \cite{Long2010}),
and MCMC is a time-consuming part of Boltzmann machine learning.
Reducing the computational cost for Boltzmann sampling will substantially
speed up various machine learning tasks \cite{Hinton2002}.

Recently there have been extensive efforts to tackle such hard computational tasks
by building physical systems that can solve a given problem by using their own physical dynamics.
Finding the ground state of the Ising model is one of the main focuses of physical computing.
Various types of implementation have been proposed for Ising-type optimization problems,
such as a superconducting qubit-based quantum annealing machine \cite{Johnson2011},
a CMOS-based annealing machine \cite{Yamaoka2015},
and a coherent optical system using network of lasers \cite{Utsunomiya2011} and
optical parametric oscillators \cite{Marandi2014,Inagaki2016}.
There is also an increasing interest in using these devices for Boltzmann sampling \cite{Dupret1996,Denil2011,Dumoulin2014} 

In this work, we propose the physical implementation of a Boltzmann sampler
for the classical XY model by using a laser network \cite{Utsunomiya2016}.
The XY model is a fundamental spin model in which spins have a continuous direction in a two-dimensional plane.
It describes interesting two-dimensional phenomena such as the Berezinskii-Kosterlitz-Thouless transition in a two-dimensional lattice \cite{Berezinskii1971, Kosterlitz1973}.
There have been recent efforts to build a XY model simulator by using optical systems such as a coupled laser system \cite{Nixon2013} and a coupled polariton system \cite{Berloff2016}.

There is a relationship between the XY model and the complex-valued neural network \cite{Zemel1995}.
Efficient sampling for the XY model has potential application training neural networks \cite{Reichert2014}.

The dynamical behavior of XY spins is also known as the Kuramoto model \cite{Acebron2005} in the field of dynamical system theory.
Simulating the dynamics of XY spins is also of great importance to understand
synchronization phenomena in complex networks \cite{Rodrigues2016}.
Directly observing the dynamics toward synchronization in laser networks 
may also be applied to computationally difficult task such as community detection \cite{Arenas2006}.
Our objective is to have our laser implementation of the XY model pave the way towards physical computation with continuous variables.

\begin{figure}
  \centering
  \includegraphics[width=230pt]{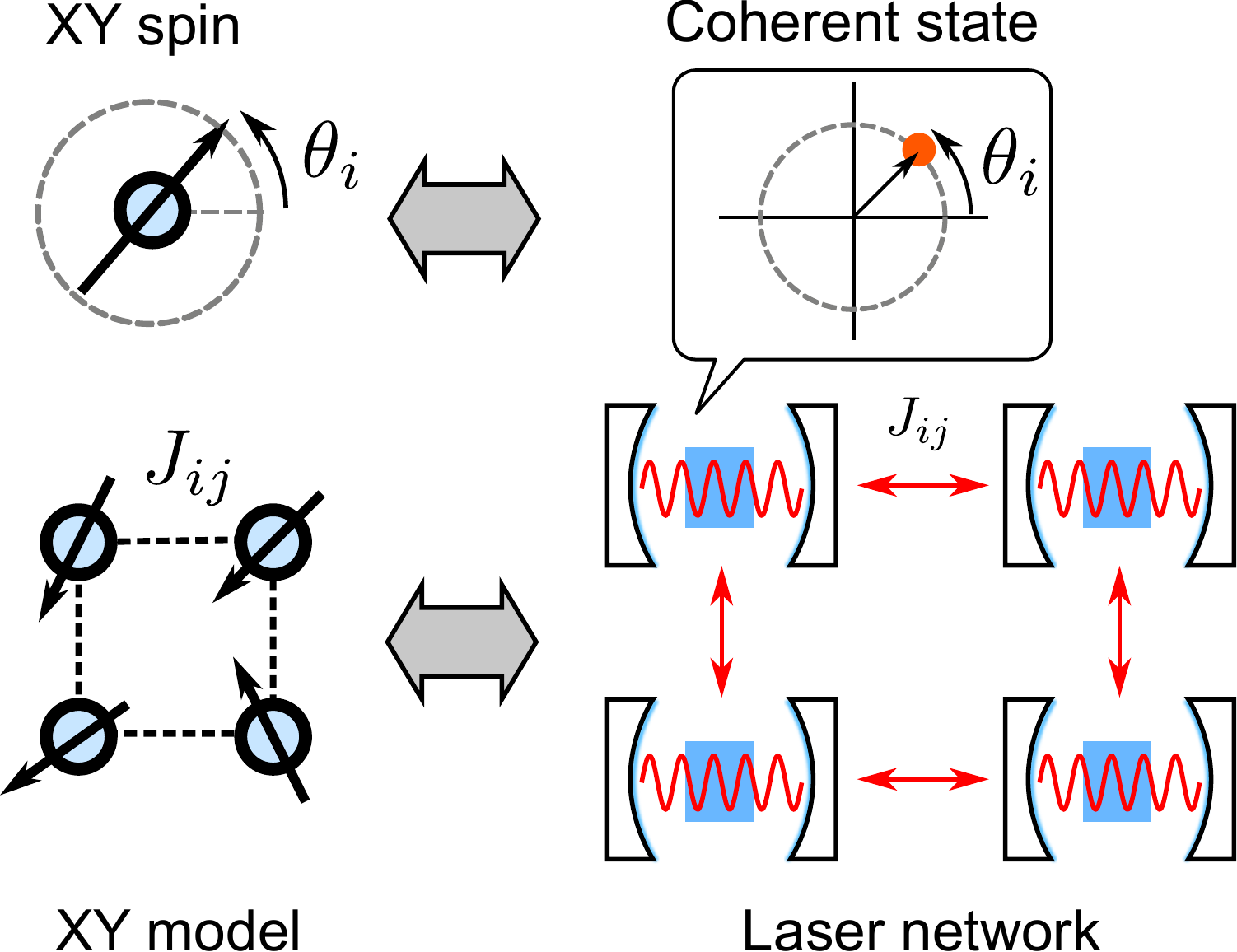}
  \caption{Mapping of the XY model onto the laser network. The angle of XY spin corresponds to the phase of coherent state generated in a laser cavity.
  The interaction between XY spins can be implemented by mutual injection between lasers.}
  \label{fig:XY_laser_mapping}
\end{figure}
Our approach to implementing the XY model with a laser network is based on the relationship between coupled lasers and the Kuramoto model \cite{Oliva2001, Acebron2005, Nixon2013}.
The mapping of the XY model onto the laser network is shown in Fig.~\ref{fig:XY_laser_mapping}.
When a laser is operated well above threshold, the phase of the laser has a $U(1)$ degree of freedom. We map the angle of the XY spin onto the phase of the laser.
The interaction between the XY spins is implemented by the mutual coupling between
lasers in the laser network.

Here, we describe how the laser network obtains a sample from the Boltzmann distribution of the XY model.
Suppose that $N$ lasers with the same wavelength are coupled to each other.
The equations of motion for such coupled lasers are described as the following
Langevin equations under the adiabatic elimination of atomic degrees of freedom \cite{Oliva2001}:
\begin{align}
  \frac{\dd A_i}{\dd t} = \frac{1}{2}\left[
    g(A_i) - \gamma\sub{c}
  \right]A_i
  + \frac{\gamma\sub{inj}}{2}\sum_{\{j:j\neq i\}} J_{ij}A_j + \xi_i(t), \label{eq:1}
\end{align}
where we denote the slowly varying amplitude of the $i$-th laser field as $A_i(t)$.
The gain function is given by $g(A) = g_0/(1 + |A|^2/n_0)$ with small-signal gain $g_0$
and saturation photon number $n_0$.
The cavity decay rate is given by $\gamma\sub{c}$ and the mutual injection rate between lasers is denoted as $\gamma\sub{inj}$.
The connections between lasers is represented as a matrix $J$ with entries $J_{ij}$.
We assume that the amplitude noise of this system is given by complex white noise $\xi_i(t)$ with diffusion rate $D$, that is,
$\langle \xi_i(t)\xi_j^*(t') \rangle = 2D\delta_{ij}\delta(t - t'),\ \langle \xi_i(t)\xi_j(t') \rangle = 0$. The diffusion coefficient is given by $D = \gamma\sub{c}/2$ for the intrinsic quantum fluctuation.

The potential function for the Langevin equation is
\begin{align}
  \tilde{H}(\vct{A}) &= -\frac{1}{2}\sum_i \left[
    g_0 n_0 \ln(n_0 + |A_i|^2) - \gamma\sub{c}|A_i|^2
  \right] \nn
   &\hspace{10pt} - \frac{\gamma\sub{inj}}{2}\sum_{i, j} J_{ij}A_i^*A_j  \label{eq:2}
\end{align}
and the Langevin equation can be written as
$\dd A_i / \dd t = - \partial \tilde{H} / \partial A_i^* + \xi_i(t)$.
Assuming the connection matrix $J$ is Hermitian, the potential function
$\tilde{H}(\bm{A})$ becomes real-valued.
For such a case, the steady-state distribution of the laser amplitudes can be expressed as $P_\mathrm{st}(\bm{A}) \propto \exp ( -\tilde{H}(\bm{A})/D )$
\cite{Risken1974, Gordon2002}.

 We further assume that the injection terms are small such that each laser is stabilized independently at the steady-state photon number,
 so the steady-state distribution can be approximated as
\begin{equation}
  P_\mathrm{st}(\bm{A}) \propto \prod_i \delta(|A_i|^2 - n_\mathrm{s}) \exp\left[n_\mathrm{s}\frac{\gamma\sub{inj}}{D} \sum_{i<j} J_{ij} \cos(\theta_i - \theta_j) \right],  \label{eq:3}
\end{equation}
where $n\sub{s} = (g_0 - \gamma\sub{c})n_0/\gamma\sub{c}$ is the steady-state
average photon number for each laser.
Equation~\eqref{eq:3} shows that the steady-state distribution of the phases of the laser network obeys the Boltzmann distribution of the XY Hamiltonian:
\begin{equation}
  H(\bm{\theta}) = - \sum_{i < j} J_{ij}\cos(\theta_i - \theta_j).  \label{eq:4}
\end{equation}
The effective inverse temperature $\beta$ is given by
\begin{equation}
  \beta = n_\mathrm{s}\frac{\gamma\sub{inj}}{D} = \frac{\gamma\sub{inj}}{D_\theta},  \label{eq:5} 
\end{equation}
where $D_\theta = D / n\sub{s}$ is the Schawlow-Townes diffusion constant for the phase variable.

\begin{figure}
  \centering
  \vspace{10pt}
  \includegraphics[width=240pt]{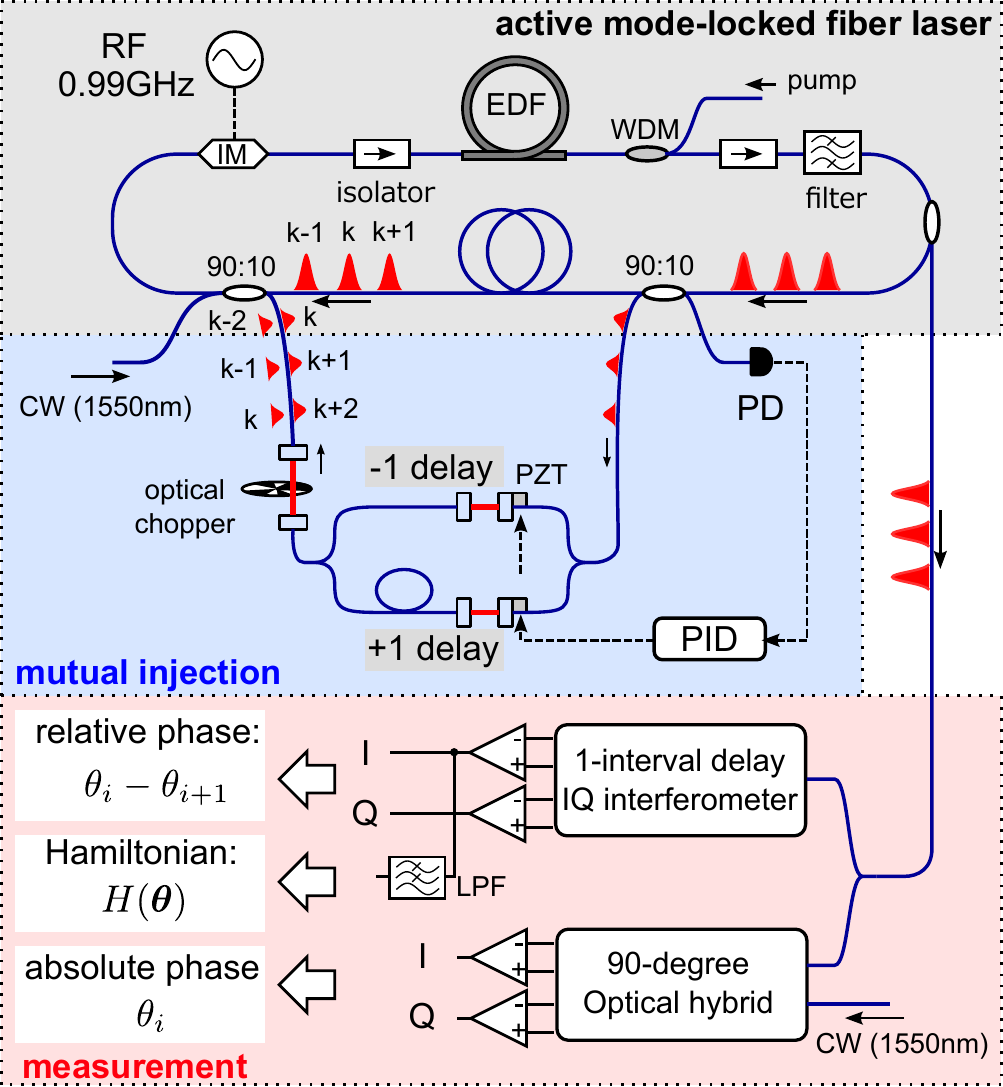}
  \caption{Schematic of experimental setup. The actively mode-locked laser (PriTel) is customized to have the two extra 90/10 couplers for injection. The lengths of delay lines can be adjusted in the free space parts.
  The absolute phase of each pulse is measured by a $90^{\circ}$ optical hybrid (Kylia, COH24-X) followed by balanced photodetector (Finisar, BPDV2150RM).
  The energy of the ferromagnetic XY ring is directly measured by a one-interval delay IQ interferometer (see supplemental material).
  The phase of the $\pm 1$ delay lines are stabilized by the external continuous-wave (CW) laser (Koshin-Kogaku, LS-601A), whose wavelength is tuned to be inside the spectrum of the mode-locked laser.
  The same CW laser is used as the LO signal for the absolute phase measurement.}
  \label{fig:experimental_setup}
\end{figure}

As a simple demonstration of a Boltzmann sampler, we constructed a one-dimensional ring of XY spins with identical ferromagnetic coupling
and experimentally observed the ground state and the winding excited states which are low-energy excitations in this system.
The experimental setup is shown in Fig~\ref{fig:experimental_setup}.
We used an actively mode-locked fiber laser, and each pulse in the fiber cavity was regarded as an independent XY spin.
The connections between adjacent pulses are implemented by using $\pm 1$-interval
optical delay lines.

The mode-locked Er-doped fiber laser has a center wavelength of $1550\,\U{nm}$, and a repetition rate of $0.99\,\U{GHz}$, and the number of pulses inside the cavity was $N=100$.
The pulse duration was measured to be $3\,\U{ps}$ and intra-cavity optical power was estimated to be $2.5\,\U{mW}$.
The 90/10 coupler placed in the fiber cavity picks up the portion of light from each pulse, and the following 90/10 coupler injects it back into the forward and backward adjacent pulses through the $\pm 1$ interval delay lines, respectively.
The injection ratio can be varied by tuning the coupling ratio of
two collimators in the middle of the delay lines.
The phases of injection was stabilized to be in-phase (ferromagnetic) by using an external continuous-wave (CW) laser.
These two delay lines can also be switched on and off simultaneously using an optical chopper.
The absolute phase of each pulse was measured on the basis of interference with the external CW laser by using a $90^{\circ}$ optical hybrid followed by pairs of balanced photodetectors (see supplemental material).
We also measured the relative phase of adjacent pulses by using a 1-interval delay in-phase/quadrature-phase (IQ) interferometer.
The in-phase components of adjacent pulse interference is given by
$\Re[A_iA_{i+1}^*] \simeq n\sub{s}\cos(\theta_i - \theta_{i+1})$.
The low-pass filters (cut-off frequency: $1.9\, \U{MHz}$) placed after the balanced photodetector add up the cosine components during about five round trips,
and the output signal directly corresponds to the energy of the one-dimensional XY Hamiltonian.

\begin{figure}
  \centering
  \includegraphics[width=250pt]{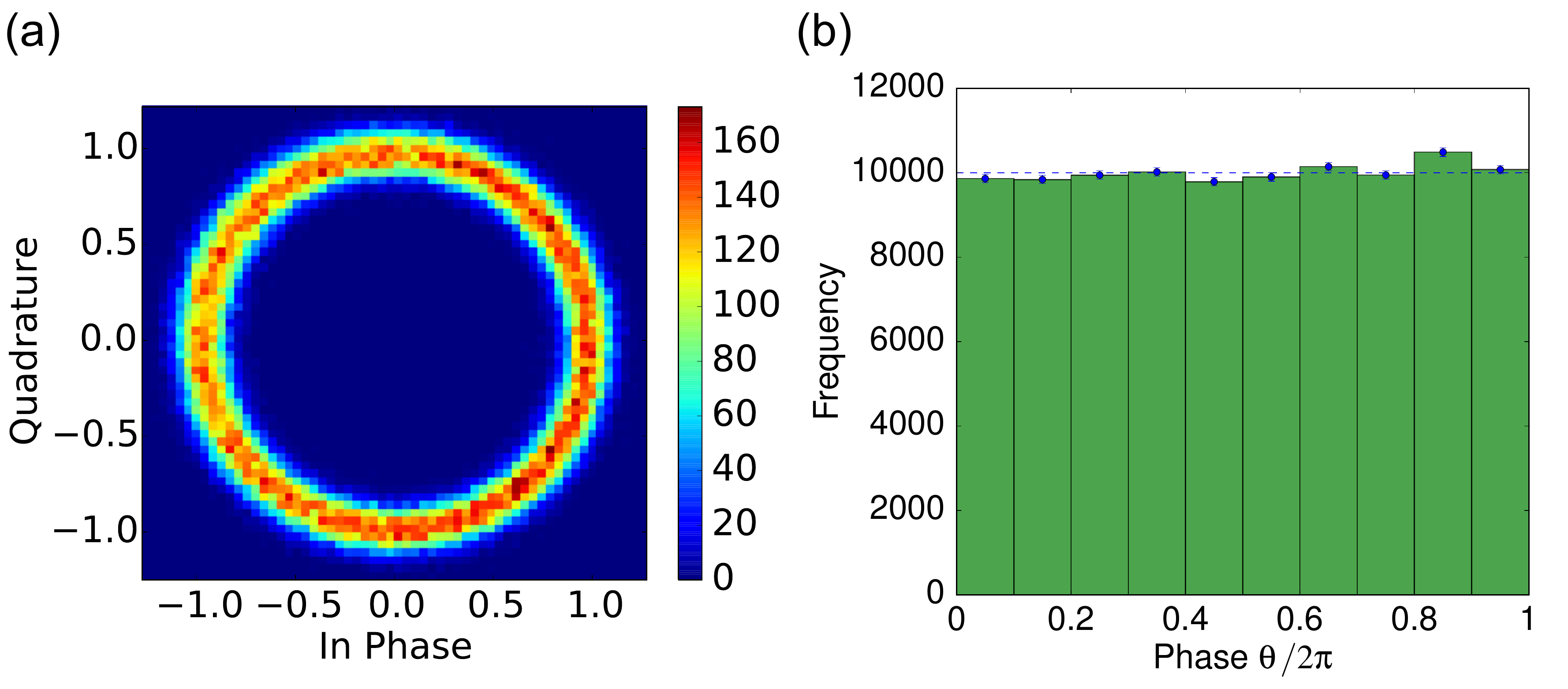}
  \caption{Histogram of the IQ measurement generated by 100 runs of mode-locked laser.
  (a) The distribution of phases of $10,000$ pulses are plotted on the IQ plane. (b) The uniform distribution of $10$ bins between $0$ to $1$ generated by nomalizing the phase $\theta$ of 100,000 pulses.}
  \label{fig:IQ_hist}
\end{figure}

We first confirmed the independence of the phases of $100$ uncoupled pulses in our mode-locked fiber laser.
The optical paths for injection was blocked, and we measured the distribution of the relative phase between adjacent pulses.
The distribution of the relative phase measurements of 1,000 runs (a total of 100,000 pulses) is shown in Fig.~\ref{fig:IQ_hist}~(a).
To confirm the uniformity of the phase distribution, we normalized the angle $\theta$ into $[0, 1)$ and plotted the histogram of the angle measured in $10$ bins,
as shown in Fig.~\ref{fig:IQ_hist}~(b).
The measured distribution is close to the uniform distribution.

\begin{figure}
  \centering
  \includegraphics[width=240pt]{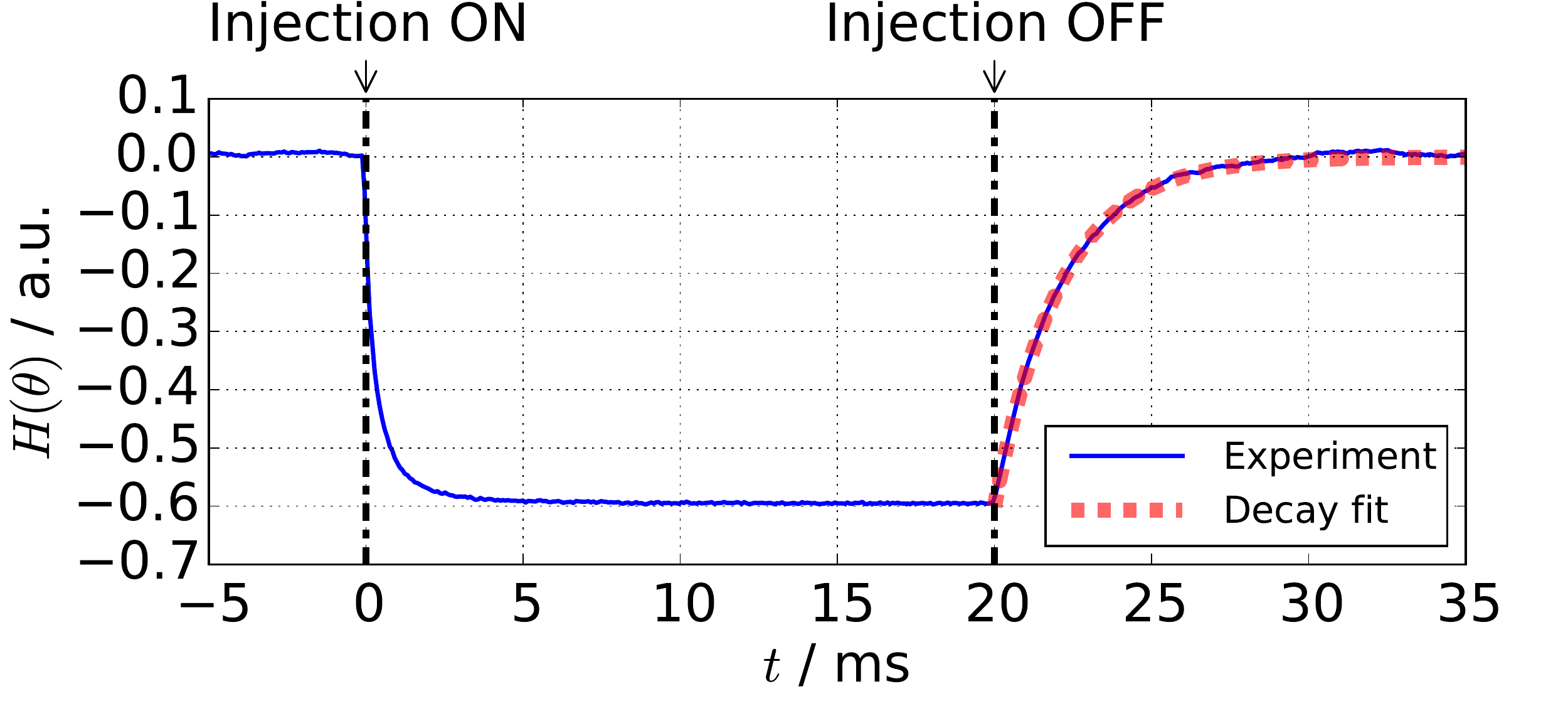}
  \caption{The time evolution of the energy of the XY ring. The solid blue line shows
  the averaged readout of 1-interval delay measurement, which corresponds to
  the energy of the XY ring. The dotted red line starting from $t=20\,\U{ms}$ is the fitting curve obtained by an exponentially decayed function.
  The injection was turned on at $t=0\,\U{ms}$ and turned off at $t=20\,\U{ms}$.
  The rise and fall times respectively correspond to the computational and
  diffusion times of the laser system.
  }
  \label{fig:decay_fit}
\end{figure}

We next introduced the $\pm 1$ bit delay lines and measured the time to
reach the steady state.
The Hamiltonian of the one-dimensional ferromagnetic XY model is given by
\begin{equation}
H(\vct{\theta}) = - \sum_{i=1}^N \cos (\theta_{i} - \theta_{i+1}),  \label{eq:6}
\end{equation}
where $\theta_{N+1} = \theta_1$ due to periodic boundary condition.
The laser dynamics were studied by turning on/off the injection path
with the optical chopper.
The rotation frequency of the optical chopper was set to $25\,\U{Hz}$.
The coupling ratio of each delay line was set to $4.2 \times 10^{-7}$.
Figure~\ref{fig:decay_fit} shows the time evolution of the energy of the XY ring measured by the one-interval-delay interference.
The blue line in Fig.~\ref{fig:decay_fit} shows the readout of the
one-interval delay measurement averaged over $120$ trials.
At $t=0\,\U{ms}$ the optical chopper was opened and the injection was turned on,
and at $t=20\,\U{ms}$ the injection was turned off.
The energy of the XY ring suddenly decreased once the injection was turned on
and gradually come back to 0 after turning off the injection.
The time to reach $80\%$ of the final energy was $1\,\U{ms}$.

We can estimate the phase diffusion coefficient $D_\theta$ from the
decay time of the ferromagnetic order after turning off the injection.
The phase diffusion of the laser obeys the Langevin equation:
$\dd \theta\sub{i} = \sqrt{D_\theta} \dd W_i$. Thus,
the averaged dynamics of the energy without injection can be calculated as
$\bracket{H(\vct{\theta}(t))} = \exp(-D_\theta t)\bracket{H(\vct{\theta}(0))}$.
The exponential fit of the energy decay is shown in Fig.~\ref{fig:decay_fit} as the dotted red line.
We obtained the experimental value of the phase diffusion coefficient as
$D_\theta = 0.480 \pm 0.002 \,\U{kHz}$ from the fitting parameter.
The Schawlow-Townes limit of the phase diffusion constant is estimated to be of the order of Hz in our system. Thus the phase diffusion coefficient is dominated by technical noise.

Finally, we observed the sampled distribution of the XY spin states for $1,000$ runs.
We set the coupling ratio of each delay line to be $1.3 \times 10^{-5}$. When we tuned the coupling ratio to more than twice this value, the oscillation of the mode-locked laser itself became unstable.
The corresponding injection rate for Eq.~\eqref{eq:1} was calculated as $\gamma\sub{inj} = 72\,\U{kHz}$.
From Eq.~\eqref{eq:5}, the expected temperature of the simulated XY model is $\beta = 150$.
In this experiment, we rotate the optical chopper with the frequency of $50\,\U{Hz}$ and sampled the phase distribution at $t = 5\,\U{ms}$.

The one-dimensional XY model is known to have winding states as local minima of the XY Hamiltonian.
The winding state with the winding number $m$ is $\theta_{k} = \theta_0 + 2\pi mk/N$.
This state corresponds to the situation where the spins are rotated slowly along
the connected direction and finally are rotated by a total of $2 m \pi$ after one round trip of the one-dimensional ring.
The energy of this state is $E_m = - N\cos(2\pi m / N)$.

\begin{figure}[t]
  \centering
  \includegraphics[width=240pt]{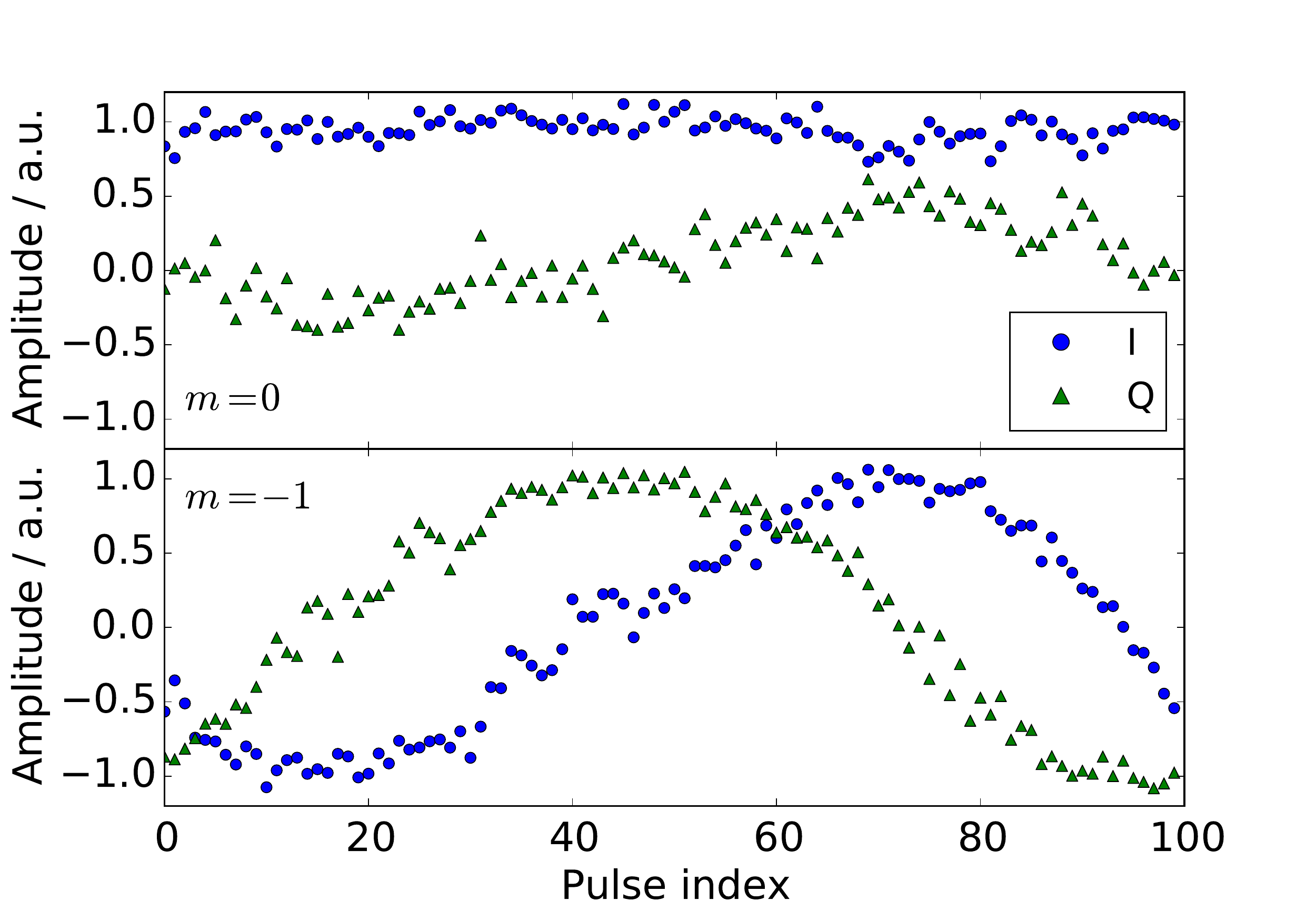}
  \caption{Observed amplitudes of 100 pulses after $5\, \U{ms}$ ferromagnetic injection. The upper and lower panels correspond to the $m=0$ and $m=-1$ winding states, respectively. The blue circles and green triangles show the in-phase and quadrature components of normalized amplitudes.}
  \label{fig:IQ_amp_ferro}
\end{figure}

The states of the XY spins we sampled during 1,000 runs were mostly one of these winding states due to the low effective temperature.
Two typical states observed in the absolute phase measurement are shown in Fig.~\ref{fig:IQ_amp_ferro},
where the upper and lower panel of Fig.~\ref{fig:IQ_amp_ferro} are the observed states corresponding to winding numbers $m = 0$ and $m = -1$, respectively.

\begin{figure}[t]
  \centering
  \includegraphics[width=250pt]{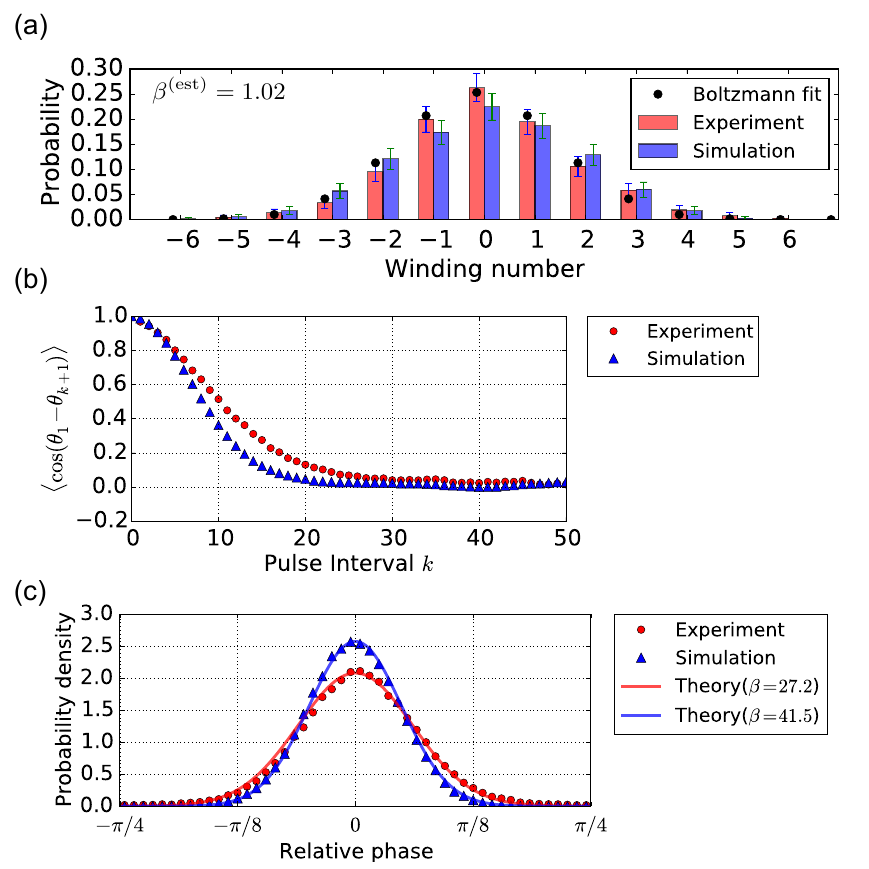}
  \caption{Comparison of the experiments and numerical simulation results. (a) The winding number distribution. The red (left) and blue (right) bars show the results of experiment and numerical simulation, respectively. The black circles are the fitting of the experimental results with the Boltzmann distribution. (b) The correlation function. The red circles and blue triangles show the results of experiment and numerical simulation, respectively. The solid red and blue lines show the theoretical fitting for those results.
  (c) The probability density of relative phase. The red circles and blue triangles show the results of experiment and numerical simulation, respectively. The solid red and blue lines show the theoretical fitting for those results.
  (See supplemental material for the theoretical expressions of the fitting curves in (b) and (c))}
  \label{fig:ferro_results}
\end{figure}

The observed winding number distribution for 1,000 runs is shown in Fig.~\ref{fig:ferro_results}~(a) as the red bars.
From this distribution, we estimated the temperature of realized
distribution by fitting the distribution with $P(m) \propto \exp(-\beta\sur{(est)} E_m)$.
The estimated temperature was $\beta\sur{(est)} = 1.02\pm 0.04$.
We also numerically simulated the Langevin dynamics of Eq.~\eqref{eq:1} and compared the results with the experimentally observed distribution.
In the numerical simulation, we used the following parameters: $\gamma\sub{c} = 50 \,\U{MHz}$, $g_0 = 100\,\U{MHz}$, $n_0 = 1.0 \times 10^7$, $\gamma\sub{inj} = 75\,\U{kHz}$, and $D_\theta = 0.5\,\U{kHz}$. The values of the parameters were estimated from the experimental setup.
The winding number distribution from the numerical simulation is shown as the blue bars in Fig.~\ref{fig:ferro_results}, which agree well with the experimental results.
The estimated temperature for the numerical simulation is $\beta\sur{(est)} = 0.75 \pm 0.02$.

We also compare the experimental and numerical simulation results for
the correlation function and relative phase distribution of the adjacent pulses,
as shown in Fig.~\ref{fig:ferro_results}~(b) and (c).
We can confirm that the experimental results agree well with the numerical simulation of the Langevin dynamics.
The theoretical fit of the result of relative phase distribution indicates the effective temperature of $\beta = 27.2 \pm 0.2$. Thus, the laser system is locally well-thermalized compared to the winding number distribution, which is a global feature of the XY system.
(See supplemental material for numerical analysis of global equilibration)

In conclusion, we have proposed and demonstrated a way to implement Boltzmann sampling for the XY model by using a mode-locked fiber laser.
Since the mode-locked fiber laser has a small phase diffusion coefficient,
we can achieve an extremely low effective temperature of the simulated XY model with a small injection ratio.
We experimentally observed that the laser system correctly found the local minima of the one-dimensional XY Hamiltonian.
We confirmed that the laser system is locally well thermalized, and the realized distribution agrees well with a numerical simulation of the Langevin dynamics.

We hope that our optical implementation of the XY model will accelerate the sampling of the XY model and open up new applications for Langevin dynamics in the field of statistical physics as well as machine learning.

The authors thank Hiroki Takesue, Takahiro Inagaki, Ryan Hamerly, Kenta Takata, Yoshitaka Haribara, Hiromasa Sakaguchi, and Yutaka Takeda for valuable discussions.
This work was funded by the Impulsing Paradigm Change through Disruptive Technologies (ImPACT) Program of the Council of Science, Technology and Innovation, Japan.

\pagebreak
\widetext
\begin{center}
\textbf{\large Supplemental Materials: Simulating the classical XY model with a laser network}
\end{center}

\setcounter{equation}{0}
\setcounter{figure}{0}
\setcounter{table}{0}
\setcounter{page}{1}
\makeatletter
\renewcommand{\theequation}{S\arabic{equation}}
\renewcommand{\thefigure}{S\arabic{figure}}
\renewcommand{\bibnumfmt}[1]{[S#1]}
\renewcommand{\citenumfont}[1]{S#1}

\section{Phase measurements}

\subsection{Absolute phase measurements with independent laser}

In our experiments, the phases of laser pulses are measured via the interference with
the independent CW laser as a local oscillator.
We describe how to estimate the absolute phase of each pulse from the interference with the independent CW laser.

Let $\omega\sub{s}$ denote one of the cavity mode frequencies of the mode-locked laser.
The amplitude $\tilde{A}_k(t)$ of the $k$-th pulse is denoted by
\begin{align}
  \tilde{A}_k(t) = A_k(t)\ee^{\ii \omega\sub{s}(t + \tau\sub{R}k/N)},  \label{eq:1}
\end{align}
where $A_k(t)$ is the slowly varying amplitude, $\tau\sub{R}$ is the cavity round-trip time, and $N$ is the number of pulses inside the cavity.

Let $\omega\sub{LO}$ denote the angular frequency of the CW laser.
The wavelength of the CW laser is variable 
and we adjusted the wavelength so as to be overlapped with the spectrum of the mode-locked fiber laser.
Since the frequency comb of the mode-locked fiber laser has the angular frequencies with the spacing of $\omega\sub{cav} = 2 \pi \tau\sub{R}^{-1}$,
there exists the angular frequency $\omega\sub{s}$ that satisfies the following condition:
\begin{align}
  |\omega\sub{s} - \omega\sub{LO}| < \frac{\omega\sub{cav}}{2}.  \label{eq:2}
\end{align}

The measured amplitude $A_k\sur{(meas)}(t)$ with reference to the CW laser is written as
\begin{align}
  A_k\sur{(meas)}(t) = \tilde{A}_k(t)\ee^{-\ii \omega\sub{LO}(t + \tau\sub{R}k/N)} = A_k(t)\ee^{\ii (\omega\sub{s} - \omega\sub{LO})(t + \tau\sub{R}k/N)}.  \label{eq:3}
\end{align}
To obtain the slowly varying amplitude $A_k(t)$, we need to compensate the phase factor
coming from the frequency difference of $\omega\sub{s}$ and $\omega\sub{LO}$.

We estimated the frequency difference of $\omega\sub{s}$ and $\omega\sub{LO}$ from the results of two-round-trip data of in-phase/quadrature-phase measurements.
In our experiments, the timescale of the phase dynamics is about the order of $10\,\U{us}$ because the fastest phase dynamics is determined by $\gamma\sub{inj} = 72\,\U{kHz}$.
Since the round-trip time $\tau\sub{R} = 101\,\U{ns}$ is much shorter than the timescale of the phase dynamics, we may assume that the slowly varying amplitude $A_k(t)$ is not changed significantly after one round trip of the fiber cavity:
\begin{align}
  A_k(t + \tau\sub{R}) \simeq A_k(t)  \label{eq:4}
\end{align}
Thus, the measured amplitude after one round trip can be written as
\begin{align}
  A_k\sur{(meas)}(t + \tau\sub{R}) \simeq A_k(t)\ee^{\ii(\omega\sub{s} - \omega\sub{LO})t}\ee^{\ii (1 + k/N) \Delta \phi}  \label{eq:5}
\end{align}
where $\Delta\phi := (\omega\sub{s} - \omega\sub{LO})\tau\sub{R}$.
From Eq.~\eqref{eq:2}, the phase difference after one-round trip satisfies
$|\Delta \phi| < \pi$.

We can estimate the phase difference $\Delta \phi$ by comparing the measured amplitude over two round trips:
\begin{align}
  A_k\sur{(meas)}(t + \tau\sub{R})A_k\sur{(meas)}(t)^* = |A_k(t)|^2 \ee^{\ii \Delta\phi}  \label{eq:6}
\end{align}
Taking the angle of the summation of the inner products gives us the estimated value as
\begin{align}
  \Delta \phi\sur{(est)} &:= \mathrm{arg}
  \left[
    \sum_{k=1}^N A_k\sur{(meas)}(t + \tau\sub{R})A_k\sur{(meas)}(t)^*
  \right] \nn
  &= \mathrm{arg} \left[
    \ee^{\ii\Delta\phi} \sum_{i=1}^N |A_k(t)|^2 \dd t
  \right]
  = \Delta \phi  \label{eq:7},
\end{align}
The absolute phases of the pulses were obtained by compensating the frequency difference of $\omega\sub{s}$ and $\omega\sub{LO}$ by using $\Delta \phi\sur{(est)}$ as
\begin{align}
  A_k(t) = A_k\sur{(meas)}(t)\ee^{-\ii \Delta \phi\sur{(est)} k / N}\ee^{-\ii \Delta \phi\sur{(est)} t/\tau\sub{R} }.  \label{eq:8}
\end{align}
Since the second phase factor $\ee^{-\ii \Delta \phi\sur{(est)} t/\tau\sub{R}}$ is common for all pulses at the same time $t$, only the first phase factor $\ee^{-\ii \Delta \phi\sur{(est)} k / N}$ was compensated in our experiments.

\subsection{Relative phase measurement of adjacent pulses}

\begin{figure}
  \centering
  \includegraphics[width=400pt]{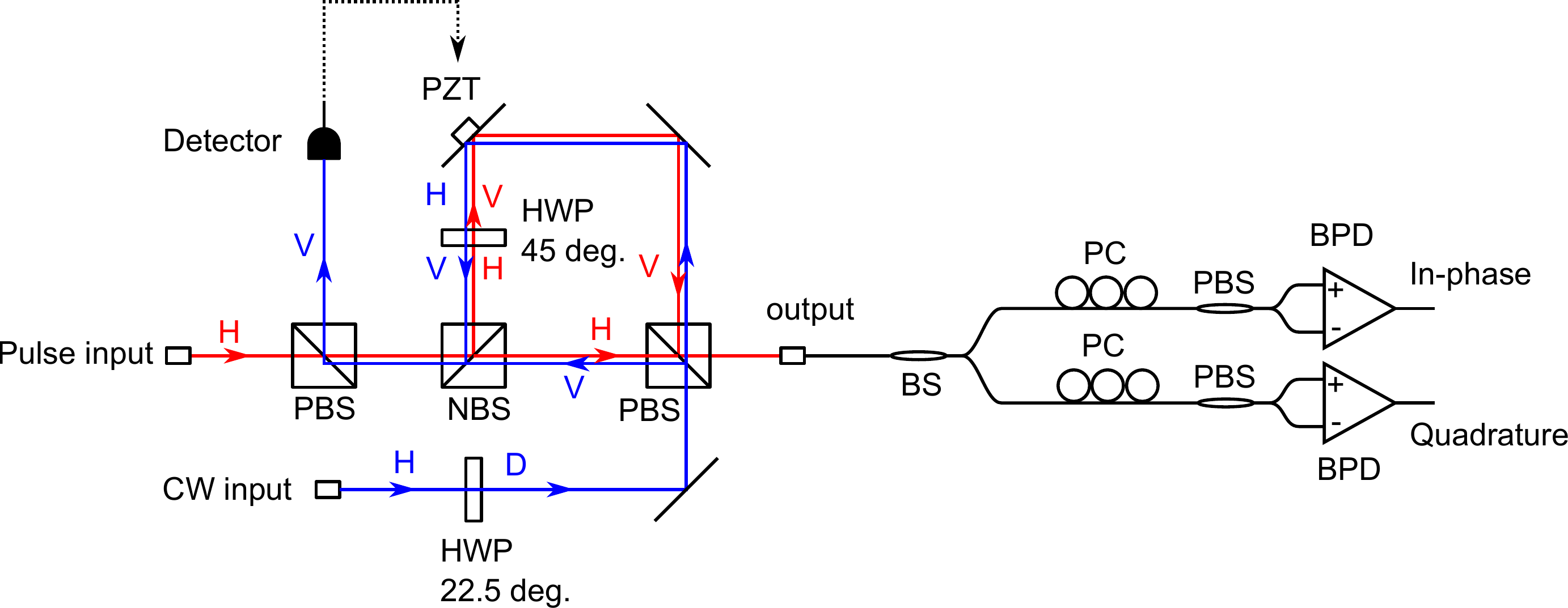}
  \caption{One-interval delay IQ interferometer for relative phase measurements.
  The input pulses (shown by the red line) are aligned so as to have the horizontal (H) polarization
  and pass through the first polarizing beam splitter (PBS). Then the pulses are
  split into two beams by the following non-polarizing beam splitter (NBS).
  The two beams travel through two interferometer arms that have different optical path lengths. The beam travelling trough the longer arm is delayed by one pulse interval, and its polarization is rotated into the vertical (V) direction with a half-wave plate.
  The two beams are recombined with the second polarizing beam splitter. Then the relative phase between adjacent pulses is mapped onto the relative phase between horizontal and vertical polarization in the single pulse.
  The converted pulses are introduced into a single-mode fiber and are split by the 50:50 coupler. The polarization of those pulses are tuned by polarization controllers (PCs), and their diagonal and circular polarization components are measured by the polarization beam splitters (PBSs) followed by the balanced photodetectors (BPDs).
  The diagonal and circular polarization components of the converted pulses correspond to the in-phase and quadrature phase components of the relative phases between adjacent pulses, respectively.
  The free-space interferometer is stabilized with the CW laser (shown by the blue lines), which is the same laser as that used for stabilizing the optical path length of $\pm 1$-interval delay lines for the mutual injection. The CW laser with diagonal polarization is inputted from the other output port of the second polarization beam splitter. The polarization of the CW laser is converted into vertical polarization in both arms and the two beams interfere when they are combined with the non-polarization beam splitter. The output beam is reflected at the first polarization beam splitter and is measured by the photodetector. The interference signal is fed back to the piezoelectric actuator driving the mirror placed in the longer arm of the free-space interferometer.}
  \label{fig:rel_phase_meas}
\end{figure}

The relative phases of adjacent pulses were measured by using a one-interval delay IQ interferometer.
The experimental setup is shown in Fig.~\ref{fig:rel_phase_meas}.

The one-interval delay IQ interferometer is composed of two parts. 
The first part consists of the free-space delay-line interferometer.
This part splits the pulses into two beams and delays one of them by one pulse interval. Furthermore, the polarization of the beam is rotated into the orthogonal polarization in the longer arm. Then two beams are recombined.
As a result, the relative phase between the two adjacent pulses is converted into the relative phase between the horizontal and vertical polarization in the single pulse.
The length of the delay line interferometer is stabilized with the CW laser which travels through the interferometer in the opposite direction to the measured pulses.

The second part is composed of fiber optics, and used for the measurements of the in-phase and quadrature phase components of the relative phase difference between horizontal and vertical polarization. 
The input pulses are coupled into two fibers with a 50:50 fiber beam splitter (BS).
One of the fiber outputs is used for the in-phase measurement and the other is used for the quadrature-phase measurement.
In each fiber, the polarization of the pulses is controlled by the polarization controllers (PCs) so as to be measured in a proper basis by the following polarizing beam splitters (PBSs) and the balanced photodetector (BPDs).

The in-phase and quadrature phase components of the one-interval delay IQ measurements are written as
\begin{align}
  I_i &= \Re[A_iA_{i+1}^*] \simeq n\sub{s}\cos(\theta_i - \theta_{i+1})  \label{eq:9} \\
  Q_i &= \Im[A_iA_{i+1}^*] \simeq n\sub{s}\sin(\theta_i - \theta_{i+1})  \label{eq:10}
\end{align}
Thus, the summation over one-round trip of the in-phase components 
is proportional to the Hamiltonian of the one-dimensional XY ring:
\begin{align}
  \sum_{i=1}^N I_i = n\sub{s}\sum_{i=1}^N \cos(\theta_i - \theta_{i+1}).  \label{eq:11}
\end{align}
In our experiments, this value was directly measured by averaging the output over about five round trips with a low-pass filter (cutoff frequency: $1.9\, \U{MHz}$).

\section{Equilibration time of one-dimensional XY model}

In our experiments, the effective temperature estimated from the winding number distribution was much different from the expected temperature determined from the
ratio between injection and diffusion.
This difference comes from the slow equilibration of the Langevin dynamics for the one-dimensional XY model.
In this section, we numerically analyze the equilibration of the winding number distribution for various inverse temperature.

To evaluate the long-term dynamics, we further simplified the Langevin dynamics of the coupled lasers so that only phases of lasers are treated as dynamical variables.
Assuming all lasers have the same steady-state photon number $n\sub{s}$, the dynamics of phases of lasers are described by the following Langevin equations:
\begin{align}
  \frac{\dd \theta_i}{\dd t} &= -\frac{\gamma\sub{inj}}{2}\frac{\partial H}{\partial \theta_i} + \sqrt{D_\theta} \eta(t),  \label{eq:12} \\
  H(\vct{\theta}) &= - \sum_{i < j} J_{ij} \cos(\theta_i - \theta_j),  \label{eq:13}
\end{align}
where $D_\theta = D/n\sub{s}$ and $\eta(t)$ is a white noise satisfying $\bracket{\eta(t)\eta(t')} = \delta(t-t')$.

In our numerical simulation, the number of spins is set as $N=100$ and the Hamiltonian~\eqref{eq:13} is chosen to be the one-dimensional ferromagnetic ring with $J_{i, i+1} = 1$ for all $i$.
The phase diffusion constant is set as $D_\theta = 0.5 \,\U{kHz}$.
The injection ratio is set as $\gamma\sub{inj} = \beta D_\theta$ depending on the inverse temperature $\beta$ to be simulated.
We used the Euler-Maruyama method to simulate Eq.~\eqref{eq:12}.
The simulated range of $\beta$ was from $\beta = 1$ to $\beta = 10$.
The numerical simulation was repeated over 1,000 runs for each $\beta$.

\begin{figure}
  \centering
  \includegraphics[width=480pt]{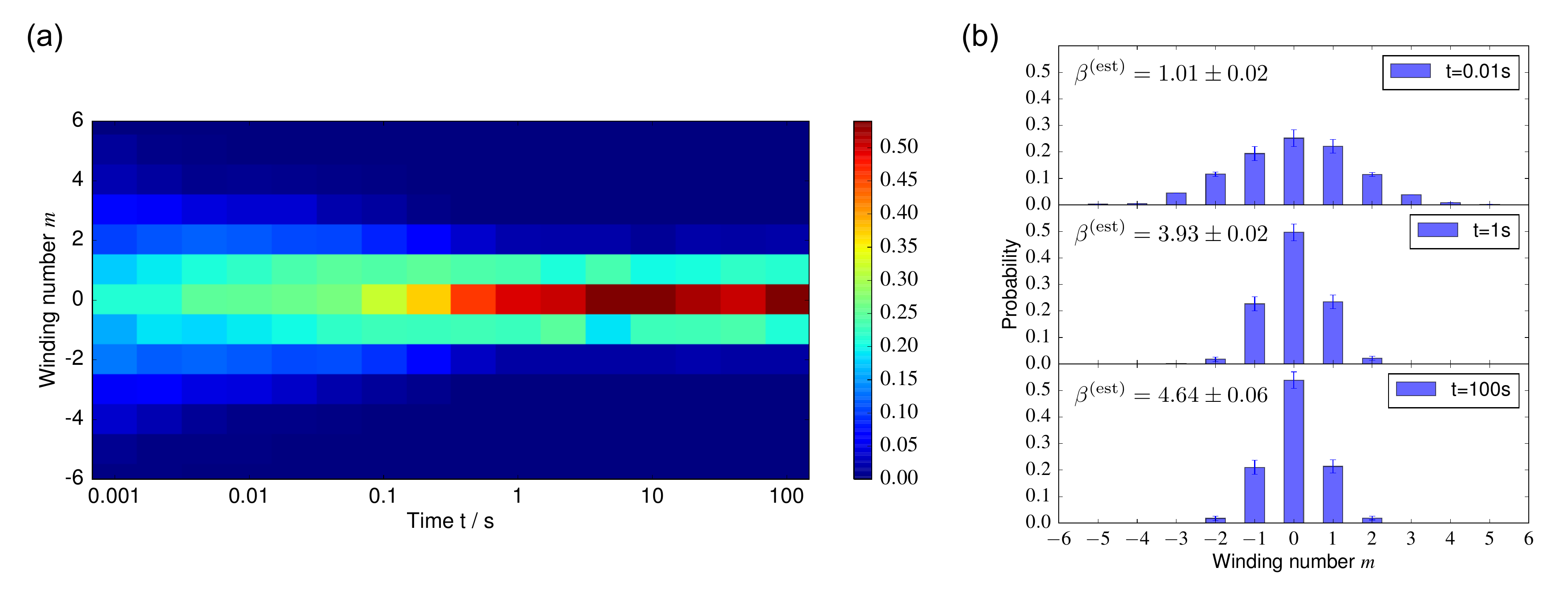}
  \caption{Time evolution of the winding number distribution for $\beta = 5$.
  (a) Probability distribution of winding number at each running time.
  The horizontal axis shows the running time $t$ plotted in a log scale.
  The color bar represents the probability of obtaining the corresponding winding number. (b) Probability distribution of winding number at $t=0.01, 1, 100 \,\U{s}$.}
  \label{fig:w_hist_time_series}
\end{figure}
\begin{figure}
  \centering
  \includegraphics[width=350pt]{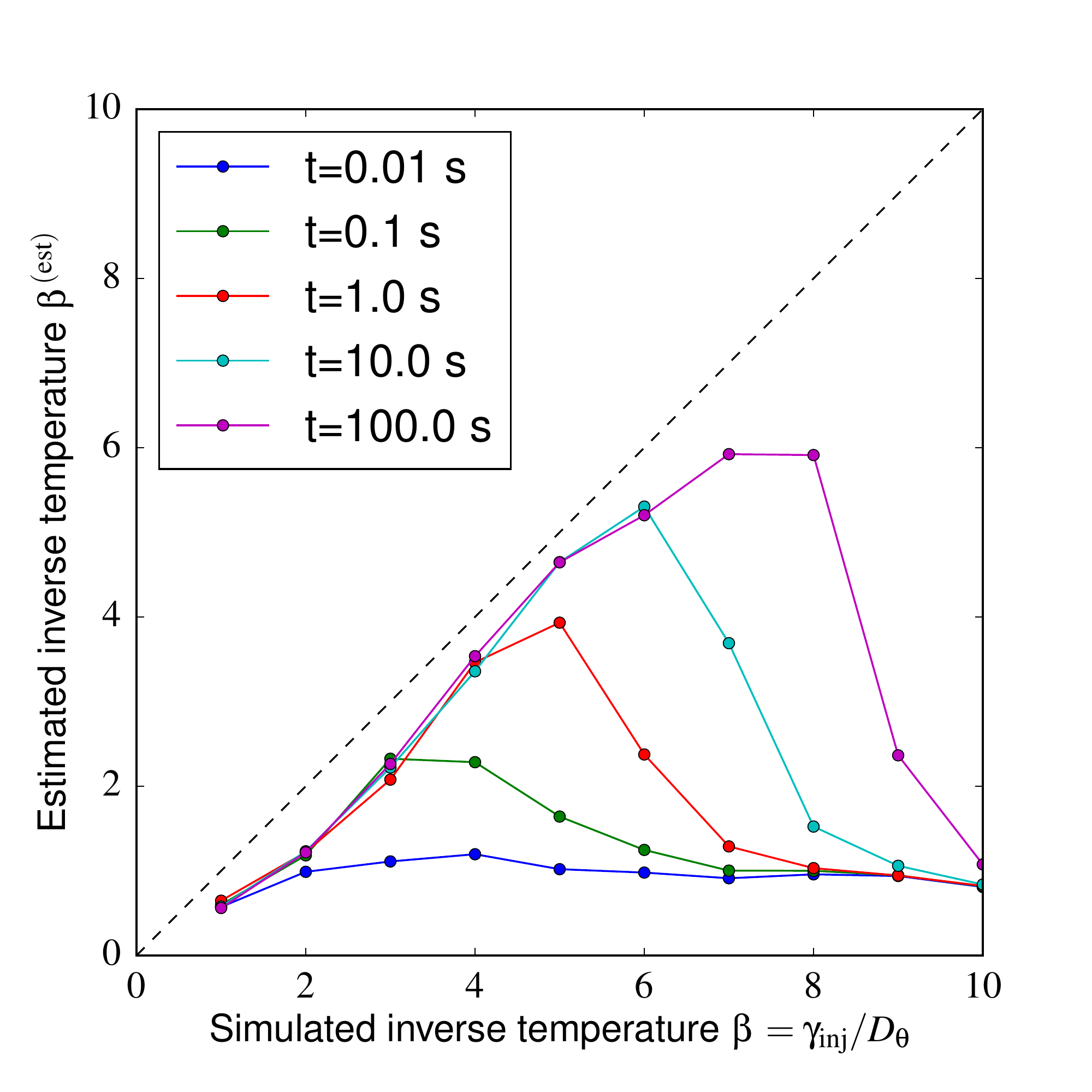}
  \caption{Running time dependence of the estimated inverse temperature.
  Each solid line represents the relationship between the simulated inverse temperature $\beta = \gamma\sub{inj}/D_\theta$ and the estimated inverse temperature for the respective running time. The dashed line represents $\beta\sur{(est)} = \beta$.}
  \label{fig:beta_vs_beta_est}
\end{figure}
Figure~\ref{fig:w_hist_time_series} shows the time evolution of the winding number distribution for $\beta = 5$, as an example.
The winding number distribution is first spread over broad range and gradually converges to a sharp distribution around $m = 0$.
The equilibration time for this case is around $t = 1\,\U{s}$.
The inverse temperature was estimated to be $\beta\sur{(est)} = 4.64 \pm 0.06$ at $t = 100 \,\U{s}$.

Figure~\ref{fig:beta_vs_beta_est} shows the estimated inverse temperature $\beta\sur{(est)}$ for various simulated inverse temperatures $\beta$ and running time $t$.
The estimated inverse temperature approaches the line $\beta\sur{(est)} = \beta$
as the running time becomes larger.
The slight deviation from the line $\beta\sur{(est)} = \beta$ possibly comes from the inaccuracy of the estimation method.
We assumed the states with winding number $m$ always have the same energy as that of the local minimum $E_m = - N \cos(2\pi m/N)$. We then estimated the inverse temperature by fitting the distribution with $p(m) \propto \exp(-\beta\sur{(est)} E_m)$.
However, the actually sampled configurations deviate slightly from the configuration of the local minimum and
the energies also deviate from the assumed one. This made the estimation slightly inaccurate.

From Fig.~\ref{fig:beta_vs_beta_est}, the distribution for larger inverse temperature $\beta$ takes a longer running time to reach the true equilibrium. The equilibration time seems to depends exponentially on $\beta$.
Even when the running time $t = 100\,\U{s}$, we can obtain the true steady-state distribution for up to around $\beta = 7$.

\section{Theory of one-dimensional XY model}

Various statistical features of the one-dimensional XY ring can be exactly calculated by using transfer matrix approach \cite{Mattis1984}.
We describe the way to calculate the partition function, the correlation function,
and the probability distribution of the relative angle between adjacent phases.

\subsection{Partition function}

The partition function $Z$ of the one-dimensional ferromagnetic XY ring is given by
\begin{align}
  Z &= \Tr[\ee^{- \beta H}],  \label{eq:14} \\
  H(\theta) &= J \sum_{i = 1}^N \cos(\theta_i - \theta_{i+1}),  \label{eq:15}
\end{align}
with the periodic boundary condition $\theta_{N+1} = \theta_1$.

Set $K = \beta J$ and define the matrix
\begin{align}
  \hat{V} = \frac{1}{2\pi}\oint \oint \dd \theta' \dd \theta \exp[K\cos(\theta' - \theta)] \ket{\theta'}\bra{\theta},  \label{eq:16}
\end{align}
where $\bracket{\theta|\theta'} = \delta(\theta - \theta')$.
The partition function of the one-dimensional ferromagnetic XY ring can be written as
\begin{align}
  Z = \Tr[\hat{V}^N].  \label{eq:17}
\end{align}
We can use the following expansion to diagonalize the matrix $\hat{V}$:
\begin{align}
  \ee^{K \cos\theta} = \sum_{n = -\infty}^{\infty} I_n(K)\ee^{\ii n \theta},  \label{eq:18}
\end{align}
where $I_n$ is the modified Bessel function of the first kind.
Define the basis vector $\ket{n}$ as
\begin{align}
   \ket{n} = \frac{1}{(2\pi)^{1/2}} \oint \ee^{\ii n \theta} \ket{\theta} \dd \theta ,  \label{eq:19}
\end{align}
then we have
\begin{align}
  \hat{V} = \sum_{n=-\infty}^{\infty} I_n(K) \ket{n}\bra{n}  \label{eq:20}
\end{align}
Thus, the partition function can be given by
\begin{align}
  Z = \sum_{n=-\infty}^{\infty} I_n(K)^N.  \label{eq:21}
\end{align}

\subsection{Correlation function}

Similar to the calculation of the partition function,
the correlation function can be given by the following form:
\begin{align}
  \bracket{\exp(\ii m (\theta_1 - \theta_{k+1}))}
  = \frac{1}{Z}\Tr [\hat{W}_m^{k} \hat{V}^{N-k}],  \label{eq:22}
\end{align}
where
\begin{align}
  \hat{W}_m = \frac{1}{2\pi}\oint \oint \dd \theta' \dd \theta \exp(\ii m (\theta' - \theta)) \exp[K\cos(\theta' - \theta)] \ket{\theta'}\bra{\theta}.  \label{eq:23}
\end{align}
and $m$ is an integer number.
Using the basis ${\ket{n}}$, the matrix $\hat{W}_m$ can be expressed as
\begin{align}
  \hat{W}_m = \sum_{n= -\infty}^{\infty} I_{n-m}(K) \ket{n}\bra{n}.  \label{eq:24}
\end{align}
Thus, we have
\begin{align}
  \bracket{\exp(\ii m (\theta_1 - \theta_{k+1}))}
  = \frac{1}{Z} \sum_{n=-\infty}^{\infty} I_{n-m}(K)^k I_n(K)^{N-k}.  \label{eq:25}
\end{align}
Since the right-hand side of the equation is real-valued, we have
\begin{align}
  \bracket{\cos(m(\theta_1 - \theta_{k+1}))} 
  = \frac{1}{Z} \sum_{n=-\infty}^{\infty} I_{n-m}(K)^k I_n(K)^{N-k}.  \label{eq:26}
\end{align}

\subsection{Probability distribution of relative angle}

The probability distribution of the relative angle between adjacent XY spins can be given as
\begin{align}
  p(\theta) &= \bracket{\delta(\theta_2 - \theta_1 - \theta)} \nn
  &= \frac{1}{Z}\frac{1}{(2\pi)^N} \oint \dd \theta_1 \cdots \oint \dd \theta_N \delta(\theta_2 - \theta_1- \theta) \prod_{i=1}^N \exp [K \cos(\theta_i - \theta_{i+1}) ] \nn
  &= \frac{\exp(K\cos\theta)}{Z} \frac{1}{(2\pi)^N} \oint \dd \theta_2 \cdots \oint \dd \theta_N 
   \exp[K\cos(\theta_N - \theta_2 + \theta)]
  \prod_{i=2}^{N-1} \exp [K \cos(\theta_i - \theta_{i+1}) ].  \label{eq:27}
\end{align}
Define the following matrix
\begin{align}
  \hat{X}(\theta) = \frac{1}{2\pi}\oint \oint \dd \theta' \dd \theta'' \exp[K\cos(\theta' - \theta'' + \theta)] \ket{\theta'}\bra{\theta''},  \label{eq:28}  
\end{align}
Then $p(\theta)$ can be written as
\begin{align}
  p(\theta) &= \frac{\exp(K\cos\theta)}{2\pi Z}\Tr[\hat{V}^{N-2}\hat{X}(\theta)]  \label{eq:29}
\end{align}
The matrix $\hat{X}(\theta)$ can be diagonalized as
\begin{align}
  \hat{X}(\theta)
   = \sum_{n=-\infty}^{\infty} I_n(K) \ee^{\ii n \theta} \ket{n}\bra{n},  \label{eq:30}
\end{align}
and then we have
\begin{align}
  p(\theta) = \frac{\exp(K\cos\theta)}{2\pi Z}\sum_{n=-\infty}^{\infty} \ee^{\ii n \theta} I_n(K)^{N-1}.  \label{eq:31}
\end{align}
Using the relationship $I_n(K) = I_{-n}(K)$, finally we obtain
\begin{align}
  p(\theta) = \frac{\exp(K\cos\theta)}{2\pi Z}\sum_{n=-\infty}^{\infty} \cos(n\theta) I_n(K)^{N-1}.  \label{eq:32}
\end{align}

\end{document}